\documentclass[aps,preprint]{revtex4}%
\usepackage{amsfonts}
\usepackage{amsmath}
\usepackage{amssymb}
\usepackage{graphicx}%
\providecommand{\U}[1]{\protect\rule{.1in}{.1in}}

\begin{document}
\title[ ]{Understanding the Planck Blackbody Spectrum and Landau Diamagnetism within
Classical Electromagnetism}
\author{Timothy H. Boyer}
\affiliation{Department of Physics, City College of the City University of New York, New
York, New York 10031}

\begin{abstract}
Electromagnetism is a \textit{relativistic} theory and one must exercise care
in coupling this theory with \textit{nonrelativistic} classical mechanics and
with \textit{nonrelativistic} classical statistical mechanics. \ Indeed
historically, both the blackbody radiation spectrum and diamagnetism within
classical theory have been misunderstood because of two crucial failures: 1)
the neglect of classical electromagnetic zero-point radiation, and 2) the use
of erroneous combinations of nonrelativistic mechanics with relativistic
electrodynamics. \ Here we review the treatment of classical blackbody
radiation, and show that the use of Lorentz-invariant classical
electromagnetic zero-point radiation can be used to explain both the Planck
blackbody spectrum and Landau diamagnetism at thermal equilibrium within
classical electromagnetic theory. \ The analysis requires that relativistic
electromagnetism is joined appropriately with simple nonrelativistic
mechanical systems which can be regarded as the zero-velocity limits of
relativistic systems, and that nonrelativistic classical statistical mechanics
is applied only in the low-frequency limit when zero-point energy makes no contribution.

\end{abstract}
\maketitle

\section{Introduction}

\subsubsection{The Rayleigh-Jeans vs the Planck Spectrum for Blackbody
Radiation}

When a physicist is asked for the theoretical difference between the
Rayleigh-Jeans spectrum and the Planck spectrum for blackbody radiation, the
response is likely to be that the first is the result of classical physics
while the second is the result of quantum physics.\cite{Mod} \ This response
may have represented the best understanding of nature in the early years of
the 20th century. \ However, it does not represent accurate physics today.
\ Today we are aware that any description of nature in terms of classical
physics must include classical electromagnetic zero-point radiation and must
recognize the demands of special relativity. \ Here we review the treatment of
classical blackbody radiation, and note that when the two missing aspects are
included accurately, then classical physics predicts not merely the
low-frequency Rayleigh-Jeans spectrum but indeed the full Planck spectrum for
thermal radiation. \ 

\subsubsection{Diamagnetism within Classical Theory}

Diamagnetism represents a second phenomenon which involves the same
misunderstandings as are involved in classical blackbody radiation. \ Current
electromagnetism textbooks provide examples of diamagnetic behavior for a
single nonrelativistic particle in a magnetic field, but then often state that
diamagnetism is not a phenomenon of classical physics.\cite{Griffiths-di}
\ The presence or absence of diamagnetism within classical physics is
controversial, partly because of a failure to distinguish what is meant by
\textquotedblleft classical physics.\textquotedblright\ \ Classical physics
includes two incompatible theories, both nonrelativistic particle mechanics as
well as relativistic classical electrodynamics. \ The Bohr-van Leeuwen theorem
applies the classical statistical mechanics of nonrelativistic particle
mechanics to the behavior of charges in an external magnetic field, and
concludes that diamagnetism does not exist.\cite{diatext} \ The Bohr-van
Leeuwen analysis does not include the magnetic energy of interacting particle
fields, since magnetic field energy is not part of nonrelativistic particle
mechanics. \ On the other hand, if one includes the magnetic field energy of
the particles (even at the level of the Darwin Lagrangian), then diamagnetism
can appear in classical physics,\cite{Essen} though the use of ideas of
nonrelativistic classical statistical mechanics may be inappropriate. \ In the
present article, we treat single-particle diamagnetism within the classical
electrodynamics of a very-low velocity particle in connecton with classical
blackbody radiation. \ The analysis gives Landau diamagnetism along with the
Planck spectrum when classical zero-point radiation is introduced and
nonrelativistic classical statistical mechanics is coupled correctly with
relativistic classical electromagnetism. \ 

\subsubsection{Coupling Nonrelativistic and Relativistic Physics in Elastic
Particle Collisions}

A sense of the erroneous results arising from the mismatch of nonrelativistic
physics and relativistic physics can be obtained by treating the elastic point
collision between two massive particles. \ The collision should be treated
relativistically using energy $U=(p^{2}c^{2}+m^{2}c^{4})^{1/2}$ and momentum
$\mathbf{p}=m\gamma\mathbf{v}$ with $\gamma=(1-v^{2}/c^{2})^{-1/2}$ for each
particle. \ If we use relativistic physics consistently, then the elastic
collision can be treated using conservation of total relativistic energy and
momentum in any inertial frame, and then, using Lorentz transformations, the
results for the collision can be transferred to any other inertial frame.
\ The results are independent of the inertial frame in which the relativistic
conservation laws were applied. \ We note that the relativistic system center
of energy $X_{C\text{of}E}=(U_{1}\mathbf{r}_{1}+U_{2}\mathbf{r}_{2}%
)/(U_{1}+U_{2}),$ where $\mathbf{r}_{1}$ and $\mathbf{r}_{2}$ refer to the
particle positions, will move with constant velocity during the collision in
every inertial frame.\cite{CofE} \ 

However, now imagine that we combine the use of \textit{nonrelativistic}
mechanics for the first particle with \textit{relativistic} mechanics for the
second. \ Thus for the first particle, we have energy $U_{1}=p_{1}^{2}%
/(2m_{1})$ and momentum $\mathbf{p}_{1}=m_{1}\mathbf{v}_{1},$ while for the
second particle, we have energy $U_{2}=(p_{2}^{2}c^{2}+m_{2}^{2}c^{4})^{1/2}$
and momentum $\mathbf{p}_{2}=m_{2}\gamma_{2}\mathbf{v}_{2}.$ \ Once again we
can use energy and momentum conservation to solve for the velocities of the
particles after the collision. \ However, the results for the final velocities
will depend upon the specific inertial frame in which energy and momentum
conservation was applied. \ 

The situation is visualized most easily for the elastic collision of particles
of equal rest mass $m=m_{1}=m_{2}.$ \ In the fully relativistic collision, the
particles exchange energy and momentum in every inertial frame, and the
relativistic center of energy moves with constant velocity despite the
collision. \ However, in the situation involving the coupling between
\textit{nonrelativistic} mechanics for the first particle and
\textit{relativistic} mechanics for the second, the use of energy and momentum
conservation will not lead to exchange of energy and momentum between the
particles. \ In the center of momentum frame where $0=\mathbf{p}%
_{1}+\mathbf{p}_{2}=m\mathbf{v}_{1}+m\gamma_{2}\mathbf{v}_{2},$ both the
center of energy and the center of mass move with non-zero velocity; the
center of energy has velocity $\mathbf{V}_{C\text{of}E}=[(mv_{1}%
^{2}/2)\mathbf{v}_{1}+(m\gamma_{2}c^{2})\mathbf{v}_{2}]/[(mv_{1}%
^{2}/2)+(m\gamma_{2}c^{2})]$ and the center of mass has velocity
$\mathbf{V}_{C\text{of}M}=(m\mathbf{v}_{1}+m\mathbf{v}_{2})/(m+m)$. \ On
collision, each particle retains its own energy, having a final velocity whose
magnitude is unchanged from its initial velocity, but the sign of the velocity
is reversed. \ In this situation, both the center of energy and the center of
mass have their velocities reversed and so do not retain their values on
collision. \ 

In fully-relativistic physics, the center-of-energy conservation law always
holds; if no external forces are present, the center of energy always moves
with constant velocity.\cite{CofE} \ In nonrelativistic mechanics, the
center-of-mass conservation law always holds; if no external forces are
present, the center of mass moves with constant velocity. \ However, in our
simple collision example, we see that if we mix relativistic and
nonrelativistic physics, then neither conservation law holds; the theory is
neither relativistic nor nonrelativistic. \ For the mixed
relativistic-nonrelativistic situation, the conservation laws of energy and
momentum can still be used, but the results will depend explicitly upon the
inertial frame in which the energy and momentum conservation laws were
applied, and the results are valid in neither nonrelativistic nor relativistic theory.

The inaccurate combing of relativistic electromagnetism with nonrelativistic
mechanical systems is involved in the erroneous ideas regarding the blackbody
spectrum and regarding diamagnetism within classical physics in the early
years of the 20th century. \ Today, the erroneous ideas are still presented in
the textbooks of modern physics.\cite{Mod} \ 

\subsubsection{Consistent Classical Derivation of the Planck Spectrum and of
Diamagnetism}

In addition to giving us a warning about the inconsistency of mixing
nonrelativistic and relativistic physics when describing nature, the situation
involving particle collisions also suggests an accurate way of using such
mixtures. \ Nonrelativistic mechanics is the low-velocity limit of
relativistic mechanics; the relativistic particle kinetic energy and momentum
go over to the nonrelativistic values in the inertial frame where a particle
has very small velocity $v/c<<1.$ \ For the two-particle collision mentioned
above, we can go to the inertial frame in which the first particle (which is
treated \textit{nonrelativistically}) has zero-initial velocity. \ Then the
motion of the second \textit{relativistic} particle will be accurately
described provided that the first particle has very small velocity also after
the collision, since at small velocity nonrelativistic mechanics represents an
accurate approximation to relativistic mechanics. \ The final velocity of the
first particle will indeed be small provided that its rest mass is taken to
the large-mass limit. \ 

This combination, involving use of the inertial frame in which a particle
always has very small velocity and also requiring that we go to the limit of
large particle mass, is a valid prescription for combing a nonrelativistic
classical mechanical particle with a relativistic classical system. \ In this
case, we can regard the nonrelativistic particle system as the valid limit of
a relativistic particle system. \ In this article, we will reexamine the
discussions of the blackbody radiation spectrum and of single-particle
diamagnetism within classical physics. We will indicate how the treatments are
modified by the inclusion of classical electromagnetic zero-point radiation
and the accurate coupling between nonrelativistic classical mechanics and
relativistic classical electromagnetic theory. \ The accurate treatments will
indeed yield correct Landau diamagnetic behavior and the Planck spectrum.

\section{Blackbody Radiation within Classical Theory}

\subsection{Scattering Calculations}

\subsubsection{Stability Under Scattering and the Importance of Relativity}

Blackbody radiation is the equilibrium spectrum of random radiation within an
enclosure whose walls are held at a constant temperature $T$. \ Now radiation
within a reflecting-walled cavity cannot bring itself to equilibrium.
\ Rather, there must be some scattering system which redistributes the energy
among the various normal modes of the cavity \ and so brings about the
spectrum of thermal equilibrium. \ Once in equilibrium, the spectrum will be
unchanged by the presence of a scatterer.

During the 20th century, there were several calculations using nonrelativistic
nonlinear scatterers to determine the theoretical equilibrium spectrum for
thermal radiation within classical theory. \ And all these nonrelativistic
nonlinear scattering calculations led to the Rayleigh-Jeans
spectrum.\cite{nonlinear} \ However, all these nonrelativistic scattering
calculations are inaccurate precisely because they attempt to couple a
nonrelativistic classical mechanical system with relativistic classical
electromagnetism.\cite{Blanco} Indeed, these nonrelativistic nonlinear
scatterers push the Lorentz-invariant spectrum of classical zero-point
radiation toward the Rayleigh-Jeans spectrum. Only fully relativistic
scatterers have the qualitative features which will leave the zero-point
radiation spectrum invariant and will allow equilibrium at the Planck
spectrum.\cite{scale} \ 

More recently there have been fully relativistic treatments of classical
thermal radiation in a relativistic accelerating frame (a Rindler frame), and
these treatments indeed lead to the Planck spectrum for thermal equilibrium
within classical physics.\cite{relbb} \ However, the analysis in a
relativistic accelerating frame involves sophistication well beyond that of
familiar nonrelativistic mechanics and basic electromagnetism. \ Therefore in
this article, we show how to combine accurately the simpler nonrelativistic
mechanics with electromagnetism.

\subsubsection{Harmonic Oscillator Systems as Relativistic Scattering Systems
for Small Velocity}

Using classical electromagnetism, we can form a one-dimensional harmonic
oscillator by trapping a bead of charge $e$ and mass $m$ on a frictionless
wire between two charges $q$ (of the same sign as $e$) which are held by
external forces at a separation $2d.$ \ Due to electrostatic forces, the
particle $e$ will undergo small oscillations at frequency $\omega
_{0}=[4eq/(md^{3})]^{1/2},$ and the forces of constraint do no work in the
inertial frame in which the charges $q$ are at rest. \ Thus in the
small-oscillation limit where the velocity of the charge $e$ vanishes
$v\rightarrow0,$ we can regard this oscillator as the limit of a relativistic
system. \ 

Indeed, a nonrelativistic classical mechanical particle in a potential will
always seek the lowest point in the potential, and small oscillations are
always harmonic oscillations. \ The nonrelativistic \textit{nonlinear}
scattering systems which were used to derive the Rayleigh-Jeans
spectrum\cite{nonlinear} can never be considered as relativistic since the
analysis depended on the nonlinear nature of the scatterer, and the scattering
associated with the nonlinearity disappears in the limit of small particle
velocity $v\rightarrow0,$ and therefore small spatial excursion$.$ \ 

Now even a charged particle undergoing strictly harmonic oscillator motion at
frequency $\omega_{0}$ with finite amplitude $x_{0}$ will radiate at all
multiples of the fundamental frequency $\omega_{0}$ with the time-average
power per unit solid angle radiated in the $n$th harmonic given
by\cite{Jackson703} \ \
\begin{equation}
\frac{dP_{n}}{d\Omega}=\frac{e^{2}c\beta^{2}}{2\pi x_{0}^{2}}n^{2}\tan
^{2}\theta\,J_{n}^{2}(n\beta\cos\theta) \label{P}%
\end{equation}
where $\beta=\omega_{0}x_{0}/c$ involves the maximum speed $v_{\max}%
=\omega_{0}x_{0}$ of the oscillating particle. \ Thus a harmonic oscillator of
finite amplitude can act as a radiation scatterer transferring energy from one
frequency to another. \ However, we notice that the power radiated into each
harmonic depends on the ratio $\beta$ of the the maximum particle speed
$v_{\max}$ to the speed of light $c$,~\ $\beta=v_{\max}/c=\omega_{0}x_{0}/c.$
\ In the nonrelativistic limit $v_{\max}/c\rightarrow0,$ all the radiation is
emitted at the fundamental frequency $\omega_{0},$ and the scatterer no longer
transfers energy from one frequency to another.

Thus the presence of a very small linear dipole harmonic oscillator within a
large reflecting-walled cavity filled with radiation will send radiation into
new directions and so tend to make the radiation isotropic,\cite{iso} but it
will not redistribute the radiation among the various frequencies. \ A linear
dipole harmonic oscillator in the small-velocity limit does not enforce any
spectrum of random radiation. \ In particular, a dipole harmonic oscillator in
the zero-velocity limit does not alter the Lorentz-invariant spectrum of
classical electromagnetic zero-point radiation. \ 

\subsection{Thermodynamic Analysis}

\subsubsection{Radiation normal Modes}

Blackbody radiation can be explored not only in terms of stability under
scattering, but also in terms of thermodynamics. \ Within classical physics,
thermal radiation corresponds to a solution of the homogeneous Maxwell
equations involving standing electromagnetic waves in an enclosure. \ Choosing
for simplicity a rectangular conducting-walled cavity of dimensions $a\times
b\times d$, the radiation inside can be written as a sum over the radiation
normal modes with vanishing scalar potential $\Phi$ and with vector potential
$\mathbf{A}$\ given by\cite{cavity}%
\begin{align}
\mathbf{A}(x,y,z,t)  &  =%
{\textstyle\sum_{l,m,n=0}^{\infty}}
{\textstyle\sum_{\lambda=1}^{2}}
q_{lmn,\lambda}c\left(  \frac{32\pi}{abd}\right)  ^{1/2}\left\{
\widehat{i}\varepsilon_{lmnx}^{(\lambda)}\cos\left(  \frac{l\pi x}{a}\right)
\sin\left(  \frac{m\pi y}{b}\right)  \sin\left(  \frac{n\pi z}{d}\right)
\right. \nonumber\\
&  +\widehat{j}\varepsilon_{lmny}^{(\lambda)}\sin\left(  \frac{l\pi x}%
{a}\right)  \cos\left(  \frac{m\pi y}{b}\right)  \sin\left(  \frac{n\pi z}%
{d}\right) \nonumber\\
&  \left.  +\widehat{k}\varepsilon_{lmnz}^{(\lambda)}\sin\left(  \frac{l\pi
x}{a}\right)  \sin\left(  \frac{m\pi y}{b}\right)  \cos\left(  \frac{n\pi
z}{d}\right)  \right\}  \label{A1}%
\end{align}
where $\widehat{\varepsilon}_{lmn}^{(\lambda)}$ with $\lambda=1,2$ are the
mutually orthogonal unit vectors satisfying $\varepsilon_{x}l+\varepsilon
_{y}m+\varepsilon_{z}n=0$ , where $q_{lmn,\lambda}$ is the time-varying
amplitude of the mode, and where the frequency of the mode is given by
$\omega_{lmn}=c\pi(l^{2}/a^{2}+m^{2}/b^{2}+n^{2}/d^{2})^{1/2},$
$l,m,n=0,1,2...$ \ The radiation energy in the box is given by $\mathcal{E}%
=[1/(8\pi)]%
{\textstyle\int}
{\textstyle\int}
{\textstyle\int}
dxdydz(E^{2}+B^{2})$ where\cite{gaussian} $\mathbf{E}=-\nabla\Phi
-(1/c)\partial\mathbf{A}/\partial t$ and $\mathbf{B}=\nabla\times\mathbf{A},$
so that\cite{cavity2}
\begin{equation}
\mathcal{E}=%
{\textstyle\sum_{l,m,n=0}^{\infty}}
{\textstyle\sum_{\lambda=1}^{2}}
(1/2)(\dot{q}_{lmn,\lambda}^{2}+\omega_{lmn,\lambda}^{2}q_{lmn,\lambda}^{2}).
\label{E}%
\end{equation}
\ Thus the energy of thermal radiation in a cavity can be expressed as a sum
over the energies of the normal modes of oscillation, with each mode taking
the form of a harmonic oscillator%
\begin{equation}
\mathcal{E}=(1/2)(\dot{q}^{2}+\omega^{2}q^{2}) \label{osc}%
\end{equation}

\subsubsection{Thermodynamics of the Simple Harmonic Oscillator}

Now the thermodynamics of a harmonic oscillator takes a particularly simple
form because the system has only two thermodynamic variables $T$ and $\omega
$.\cite{SHO} \ In thermal equilibrium with a bath, the average oscillator
energy is denoted by $U=\left\langle \mathcal{E}\right\rangle =\left\langle
J\right\rangle \omega$ (where $J$ is the action variable), and satisfies
$dQ=dU-dW$ with the entropy $S$ satisfying $dS=dQ/T.$ \ Now since $J$ is an
adiabatic invariant,\cite{Gold} the work done \textit{on} the system is given
by $dW=\left\langle J\right\rangle d\omega=(U/\omega)d\omega.$ \ Combing these
equations, we have $dS=dQ/T=[dU-(U/\omega)d\omega]/T.$ $\ $Writing the
differentials in terms of $T$ and $\omega,$ we have $dS=(\partial S/\partial
T)dT+(\partial S/\partial\omega)d\omega$ and $dU=(\partial U/\partial
T)dT+(\partial U/\partial\omega)d\omega.$ \ Therefore $\partial S/\partial
T=(\partial U/\partial T)/T$ and $\partial S/\partial\omega=[(\partial
U/\partial\omega)-(U/\omega)]/T.$ \ Now equating the mixed second partial
derivatives $\partial^{2}S/\partial T\partial\omega=\partial^{2}%
S/\partial\omega\partial T,$ we have $(\partial^{2}U/\partial\omega\partial
T)/T=-(\partial U/\partial T)/(\omega T)+(\partial^{2}U/\partial
T\partial\omega)/T+[(U/\omega)-(\partial U/\partial\omega)]/T^{2}$ or
$0=(\partial U/\partial T)/(\omega T)-[(U/\omega)-(\partial U/\partial
\omega)]/T^{2}.$ \ The general solution of this equation is\cite{SHO}%

\begin{equation}
U(\omega,T)=\omega\,f(\omega/T) \label{U}%
\end{equation}
where $\,f(\omega/T)$ is an unknown function. \ When applied to thermal
radiation, the result obtained here, purely from thermodynamics, corresponds
to the familiar Wien displacement law of classical physics.

\subsubsection{Classical Zero-Point Radiation}

The energy expression (\ref{U}) for an electromagnetic radiation mode (or for
a harmonic oscillator) in thermal equilibrium allows two limits which make the
energy independent from one of its two thermodynamics variables. \ When the
temperature $T$ becomes very large, $T>>\omega,$ so that the argument of the
function $\,f(\omega/T)$ is small, the average energy $U$\ of the mode becomes
independent of $\omega$ provided $f(\omega/T)\rightarrow const_{1}\times
T/\omega$ so that%

\begin{equation}
U(\omega,T)=\omega\,f(\omega/T)\rightarrow\omega\times const_{1}\times
T/\omega=const_{1}\times T\text{ \ \ for \ }\omega/T<<1. \label{Uh}%
\end{equation}
This is the familiar high-temperature limit where we expect to recover the
equipartition limit giving the Rayleigh-Jeans spectrum. \ Therefore we choose
this constant as $const_{1}=k_{\mathbf{B}}$ corresponding to Boltzmann's
constant. \ With this choice, our thermal radiation now goes over to the
Rayleigh-Jeans limit for high temperature or low frequency%
\begin{equation}
U(\omega,T)=\,\omega\,f(\omega/T)\rightarrow k_{\mathbf{B}}T\text{ \ \ for
}\omega/T<<1. \label{Uh1}%
\end{equation}

In the other limit of small temperature, $T<<\omega,$ the dependence on
temperature is eliminated provided $f(\omega/T)\rightarrow const_{2},$ so
that
\begin{equation}
U(\omega,T)=\omega\,f(\omega/T)\rightarrow const_{2}\times\omega\text{ \ \ for
\ }\omega/T>>1. \label{Ul}%
\end{equation}
If the the second constant does not vanish, then there exists random,
temperature-independent radiation present in the system. \ Since we are
describing nature using classical theory, this random radiation which exists
at temperature $T=0$ is classical electromagnetic zero-point radiation. \ 

We emphasize that thermodynamics allows classical zero-point radiation within
classical physics. \ The physicists of the early 20th century were not
familiar with the idea of classical zero-point radiation, and so they assumed
that $const_{2}=0$ which excluded the possibility of classical zero-point
radiation. \ In his monograph on classical electron theory,
Lorentz\cite{Lorentz} makes the explicit assumption that there is no radiation
present at $T=0.$ \ Today, we know that the exclusion of classical zero-point
radiation is an error. \ Experiments involving the (Casimir)
forces\cite{Casimir} between two uncharged conducting parallel plates show
that valid classical electromagnetic theory must assume that there is
electromagnetic zero-point radiation.\cite{Rev1} \ By comparing theoretical
calculations with experiments, one finds that the scale constant for classical
zero-point radiation appearing in Eq. (\ref{Ul}) must take the value
$const_{2}=1.05\times10^{-34}$Joule-sec. \ However, this value corresponds to
the value of a familiar constant in physics; it corresponds to the value
$\hbar/2$ where $\hbar$ is Planck's constant. \ Thus in order to account for
the experimentally observed Casimir forces between parallel plates, the scale
of classical zero-point radiation must be such that $const_{2}=\hbar/2,$ and
for each normal mode, the average energy becomes
\begin{equation}
U(\omega,T)=\,\omega\,f(\omega/T)\rightarrow U(\omega,0)=(\hbar/2)\omega\text{
\ \ for }T\rightarrow0. \label{zpr}%
\end{equation}

At this point, we have the high-temperature and low-temperature asymptotic
limits of the function $U(\omega,T).$ \ The full blackbody radiation spectrum
represents the interpolation between these limits. \ \ In an earlier article,
we suggested that, using an entropy-related function, the Planck spectrum
could be obtained as the smoothest possible interpolation between the
high-temperature and low-temperature asymptotic forms.\cite{SHO} \ Here we
will show that nonrelativistic classical statistical mechanics, when
accurately applied, will demand exactly this same interpolation given by the
Planck formula.

\subsection{Use of Nonrelativistic Classical Statistical Mechanics}

\subsubsection{The Traditional Treatment in Modern Physics}

The discussions of the blackbody radiation spectrum within classical physics
which appear in textbooks do not involve the scattering of radiation by
nonrelativistic nonlinear oscillators, or the introduction of classical
zero-point radiation allowed by thermodynamics, but rather involve the
application of nonrelativistic classical statistical mechanics. \ Classical
statistical mechanics was developed for nonrelativistic classical particle
mechanics before the ideas of special relativity and \textit{does not allow
the idea of classical zero-point energy}. \ In this article, we will always
refer to classical statistical mechanics as \textit{nonrelativistic} classical
statistical mechanics in order to emphasize its nonrelativistic character.
\ In the textbooks of modern physics, the energy equipartition theorem of
nonrelativistic classical statistical mechanics is applied directly to each
normal mode of the classical radiation field.\cite{Mod} \ The result is the
Rayleigh-Jeans spectrum. \ However, this spectrum is unjustified since it
presents the application of a result of \textit{nonrelativistic} classical
statistical mechanics to a \textit{relativistic} radiation system. \ 

\subsubsection{Dipole Harmonic Oscillator in Thermal Radiation}

In contrast to the blackbody analysis favored by the textbooks of modern
physics, the derivation favored by Planck\cite{Planck} involved first deriving
the connection between the average energy of a harmonic oscillator of
frequency $\omega_{0}$ and the average energy per normal mode of the radiation
spectrum at frequency $\omega_{0}.$ \ Here we will repeat the traditional
calculation, because we will need the result, and because some later
calculations will proceed in analogy with it. \ 

For thermal radiation in a large enclosure, one may treat the radiation not
only as a sum over normal modes as in Eq. (\ref{A1}), but also, alternatively,
as a sum over plane waves with periodic boundary conditions. \ Thus we can
take the scalar potential $\Phi_{T}$ to vanish and the vector potential as
\begin{equation}
\mathbf{A}_{T}\mathbf{(r},t)=%
{\textstyle\sum_{\mathbf{k}}}
{\textstyle\sum_{\lambda=1}^{2}}
\frac{c}{\omega}\widehat{\epsilon}(\mathbf{k},\lambda)\left(  \frac{8\pi
U(\omega,T)}{V}\right)  ^{1/2}\sin[\mathbf{k\cdot r}-\omega t+\theta
(\mathbf{k},\lambda)] \label{A2}%
\end{equation}
where the wave vectors $\mathbf{k}$ correspond to $\mathbf{k=}\widehat{i}%
l2\pi/a+\widehat{j}m2\pi/a+\widehat{k}n2\pi/a$ with $l,m,n$ running over all
positive and negative integers, $a$ is a length such that $a^{3}=V,$ and the
two mutually-orthogonal polarization vectors $\widehat{\epsilon}%
(\mathbf{k},\lambda),\lambda=1,2,$ are orthogonal to the wave vectors
$\mathbf{k.}$ Since thermal radiation is isotropic in the inertial frame of
its container, the amplitude $[U(\omega,T)]^{1/2}$ depends only on the
frequency $\omega=c|\mathbf{k}|=ck,$ and the constants are chosen so that
$U(\omega,T)$\ is the energy per normal mode appropriate for the thermal
radiation spectrum in classical physics. \ In order to describe the randomness
of the radiation, the phases $\theta(\mathbf{k},\lambda)$ are chosen as random
variables uniformly distributed on $(0,2\pi],$ independently distributed for
each $\mathbf{k}$ and $\lambda.$

When thermal radiation falls on a small dipole harmonic oscillator, modeled as
a particle of charge $e$ and mass $m$ at the end of a small spring of spring
constant $\kappa$ located at the origin of coordinates and oriented along the
$z$-axis (so that the oscillator frequency satisfies $\omega_{0}^{2}%
=\kappa/m),$ the equation of motion becomes\cite{GriffithsSHO}%
\begin{equation}
m\ddot{z}=-m\omega_{0}^{2}z+m\tau\dddot{z}+e\mathbf{E}_{Tz}\mathbf{(}0,t)
\label{Z}%
\end{equation}
where $-m\omega_{0}^{2}z$ represents the spring restoring force, $m\tau
\dddot{z}$ is the radiation damping force with $\tau=2e^{2}/(3mc^{3}),$ and
$e\mathbf{E}_{Tz}\mathbf{(}0,t)$ is the driving force of the random radiation
with $\mathbf{E}_{T}=-\nabla\Phi_{T}-(1/c)\partial\mathbf{A}_{T}/\partial t.$ \ 

For thermal radiation as given in Eq. (\ref{A2}) and the oscillator located at
the coordinate origin, the steady-state solution of Eq.(\ref{Z}) is%
\begin{equation}
z(t)=%
{\textstyle\sum_{\mathbf{k}}}
{\textstyle\sum_{\lambda=1}^{2}}
\frac{e\epsilon_{z}(\mathbf{k},\lambda)}{m}\left(  \frac{2\pi U(\omega,T)}%
{V}\right)  ^{1/2}\left\{  \frac{\exp[-i\omega t+i\theta(\mathbf{k},\lambda
)]}{(-\omega^{2}+\omega_{0}^{2}-i\tau\omega^{3})}+cc\right\}  \label{Xt}%
\end{equation}
where "$cc$" stands for the complex conjugate of the first quantity in the
curly bracket.

In thermal radiation, the mean displacement of the oscillator and the mean
velocity are both zero $\left\langle z(t)\right\rangle =0,$ $\left\langle
\dot{z}(t)\right\rangle =0,$ but the mean squares are non-zero. \ We can find
the mean-square displacement by averaging over time or averaging over the
random phases at a fixed time. \ Since the random phases $\theta
(\mathbf{k},\lambda)$ are distributed randomly and independently for each
mode, we have the averages
\begin{equation}
\left\langle \exp\left\{  i[-\omega t+\theta(\mathbf{k},\lambda)]\right\}
\exp\left\{  i[-\omega^{\prime}t+\theta(\mathbf{k}^{\prime},\lambda^{\prime
})]\right\}  \right\rangle =0 \label{av1}%
\end{equation}
and
\begin{equation}
\left\langle \exp\left\{  i[-\omega t+\theta(\mathbf{k},\lambda)]\right\}
\exp\left\{  -i[-\omega^{\prime}t+\theta(\mathbf{k}^{\prime},\lambda^{\prime
})]\right\}  \right\rangle =\delta_{\mathbf{kk}^{\prime}}\delta_{\lambda
\lambda^{\prime}} \label{av2}%
\end{equation}
which gives%
\begin{equation}
\left\langle z^{2}\right\rangle =%
{\textstyle\sum_{\mathbf{k}}}
{\textstyle\sum_{\lambda=1}^{2}}
\epsilon_{z}^{2}(\mathbf{k},\lambda)\left(  \frac{2\pi U(\omega,T)}{V}\right)
\frac{2e^{2}}{m^{2}[(-\omega^{2}+\omega_{0}^{2})^{2}+(\tau\omega^{3})^{2}]},
\label{X2av}%
\end{equation}%
\begin{equation}
\left\langle \dot{z}^{2}\right\rangle =%
{\textstyle\sum_{\mathbf{k}}}
{\textstyle\sum_{\lambda=1}^{2}}
\epsilon_{z}^{2}(\mathbf{k},\lambda)\left(  \frac{2\pi U(\omega,T)}{V}\right)
\frac{2e^{2}\omega^{2}}{m^{2}[(-\omega^{2}+\omega_{0}^{2})^{2}+(\tau\omega
^{3})^{2}]}, \label{Xdot2av}%
\end{equation}
and the average energy of the oscillator
\begin{equation}
\left\langle \mathcal{E}(\omega_{0},T\right\rangle =%
{\textstyle\sum_{\mathbf{k}}}
{\textstyle\sum_{\lambda=1}^{2}}
\epsilon_{z}^{2}(\mathbf{k},\lambda)\left(  \frac{2\pi U(\omega,T)}{V}\right)
\frac{e^{2}(\omega^{2}+\omega_{0}^{2})}{m[(-\omega^{2}+\omega_{0}^{2}%
)^{2}+(\tau\omega^{3})^{2}]} \label{Eav}%
\end{equation}
For a large box of thermal radiation, the normal modes are very closely
spaced, and therefore the sum over normal modes can be replaced by an
integral, $%
{\textstyle\sum_{\mathbf{k}}}
\rightarrow(a/2\pi)^{3}%
{\textstyle\int}
d^{3}k,$
\begin{equation}
\left\langle \mathcal{E}(\omega_{0},T)\right\rangle =\left(  \frac{a}{2\pi
}\right)  ^{3}\int d^{3}k%
{\textstyle\sum_{\lambda=1}^{2}}
\epsilon_{z}^{2}(\mathbf{k},\lambda)\left(  \frac{2\pi U(\omega,T)}{V}\right)
\frac{e^{2}(\omega^{2}+\omega_{0}^{2})}{m[(-\omega^{2}+\omega_{0}^{2}%
)^{2}+(\tau\omega^{3})^{2}]} \label{Eav2}%
\end{equation}
which is sharply peaked at $\omega=\omega_{0}.$ \ In this case, we can
integrate over all angles so that $\epsilon_{z}^{2}(\mathbf{k},\lambda)$
contributes a factor of 1/3 for each polarization, and then approximate the
integral over $k=\omega/c$ by extending the lower limit to minus infinity,
setting $\omega=\omega_{0}$ in every term except for $(-\omega^{2}+\omega
_{0}^{2})\approx2\omega_{0}(\omega_{0}-\omega),$ and using the definite
integral%
\begin{equation}%
{\textstyle\int_{-\infty}^{\infty}}
\frac{dx}{a^{2}x^{2}+b^{2}}=\frac{\pi}{ab} \label{Ib}%
\end{equation}
to obtain for the average oscillator energy\cite{Lavenda}%
\begin{equation}
\left\langle \mathcal{E}(\omega_{0},T)\right\rangle =U(\omega_{0},T)
\label{UE}%
\end{equation}
Thus the average energy $\left\langle \mathcal{E}(\omega_{0},T)\right\rangle $
of the oscillator with resonant frequency $\omega_{0}$ is the same as the
average energy $U(\omega_{0},T)$ of the radiation normal mode at the same frequency.

In this calculation, we have coupled a nonrelativistic harmonic oscillator to
relativistic electromagnetic radiation. \ The mixture of nonrelativistic and
relativistic physics is justified as a relativistic calculation only in the
limit that the oscillator velocity goes to zero.

\subsubsection{Failure of the View from the Beginning of the 20th Century}

Now the physicists of the early 20th century did not appreciate the idea of
classical zero-point radiation nor the importance of special relativity.
\ They assumed that they could apply nonrelativistic classical statistical
mechanics to the dipole oscillator motion. \ Therefore they suggested that the
average linear oscillator energy was $\left\langle \mathcal{E}(\omega
_{0},T)\right\rangle =k_{B}T,$ and that classical physics required that the
corresponding radiation mode must have average energy $U(\omega_{0}%
,T)=k_{B}T.$ \ In other words, they arrived at the Rayleigh-Jeans spectrum
because of the use of nonrelativistic classical statistical mechanics for all
mechanical oscillators, not merely for the lowest frequency oscillators.

Nonrelativistic statistical mechanics can not tolerate the idea of zero-point
energy, because within nonrelativistic statistical mechanics, \textit{all}
random energy is thermal energy associated with temperature $T$. \ Thus when
attempting to couple relativistic classical electromagnetism with ideas of
nonrelativistic classical statistical mechanics, it is not sufficient simply
to take the limit of low velocity for the oscillator to bring compliance with
relativity; we must also require that there is no contribution from zero-point
radiation. \ Thus we can apply nonrelativistic statistical mechanics only in
the limit that the velocity of the oscillator goes to the zero-velocity limit
and also that the zero-point radiation for the oscillator is very small
compared to $k_{\mathbf{B}}T,~(1/2)\hbar\omega_{0}<<k_{\mathbf{B}}T.$ \ This
situation indeed corresponds to the low-frequency section of the blackbody spectrum.

\subsubsection{Large-Mass Limit of a Harmonic Oscillator}

According to Eqs. (\ref{zpr}) and (\ref{UE}), a mechanical oscillator in
classical zero-point radiation will acquire an average mechanical energy%
\begin{equation}
\left\langle \mathcal{E}(\omega_{0},0)\right\rangle =(1/2)\hbar\omega
_{0}=(1/2)mv_{\max}^{2} \label{E0}%
\end{equation}
where $v_{\max}$ is the maximum velocity of the oscillator. \ However, a
harmonic oscillator can be regarded as the limit of a relativistic system only
when $v_{\max}$ becomes very small, $v_{\max}<<c.$ In order to make $v_{\max}$
very small for fixed average energy $\left\langle \mathcal{E}(\omega
_{0},0)\right\rangle $, we need to take the particle mass $m$ as very large.
\ However, for mechanical systems, large mass $m$ with fixed spring constant
$\kappa$ means that the oscillation frequency $\omega_{0}=(\kappa
/m)^{1/2}\rightarrow0$ as $m\rightarrow\infty.$ Thus in the large-$m$ limit
required to fit with relativity, the mechanical zero-point energy vanishes
along with the oscillation frequency, $\left\langle \mathcal{E}(\omega
_{0},0)\right\rangle =(1/2)\hbar\omega_{0}\rightarrow0$ \ as \ $\omega
_{0}\rightarrow0$ \ for \ $m\rightarrow\infty$ and $\kappa$ \ fixed. \ In this
low-frequency limit, the zero-point energy becomes ever smaller so that for
any non-zero temperature $(1/2)\hbar\omega_{0}<<k_{B}T,$ and we recover the
Rayleigh-Jeans spectrum, which is indeed found for the low-frequency radiation modes.

\section{Single-Particle Diamagnetism in Classical Physics}

\subsection{Inclusion of a Uniform Magnetic Field}

\subsubsection{Cyclotron Motion Allows a Special Large-Mass Limit}

In order to combine simple nonrelativistic mechanical systems with
relativistic electromagnetism, we must take the large-mass limit so as to
approximate a valid relativistic mechanical system. \ At the same time, we
wish to maintain the oscillation frequency $\omega_{0}$ unchanged so as to see
the influence of the electromagnetic zero-point energy on the mechanical
system. \ Clearly we need some additional variable which can be taken large in
order to compensate for the large-mass limit. \ One system in which this
occurs is the motion of a free charged particle in a uniform magnetic field
$\mathbf{B.}$ The relativistic equation for cyclotron motion is $m\gamma
v^{2}/r=e(v/c)B.$ \ In the nonrelativistic limit of small velocity, this
becomes\cite{Griffithscy} $mv^{2}/r=e(v/c)B,$ giving the frequency of
rotation
\begin{equation}
\omega_{B}=(v/r)=eB/(mc). \label{CM}%
\end{equation}
Thus\ we can maintain the cyclotron frequency $\omega_{B}$ as constant while
increasing $m$ to reach the nonrelativistic limit provided that we increase
the magnetic field $B.$ \ For fixed total energy, the velocity $v$ of the
charge becomes smaller as the mass $m$ is increased while the frequency
$\omega_{B}$ is held constant. \ Here indeed we have a simple nonrelativistic
particle system which can be regarded as the limit of a relativistic system
while maintaining the zero-point energy contribution $(1/2)\hbar\omega_{B}$. \ 

\subsubsection{Absence of Diamagnetism in Nonrelativistic Classical
Statistical Mechanics}

Diamagnetism is an equilibrium thermodynamic condition which does not exist
within nonrelativistic classical physics when we deal with charges in an
external magnetic field but we neglect the magnetic field energy of the
charges themselves \ This is the content of the Bohr-van Leeuwen
theorem.\cite{diatext} \ Of course, electromagnetism (except for
electrostatics) does not exist within \textit{nonrelativistic} classical
physics since electromagnetism (beyond electrostatics) is a relativistic
theory. \ Although single-particle diamagnetic behavior is discussed with
examples in textbooks of classical electomagnetism, there is often a
disclaimer noting that the application of nonrelativistic classical
statistical mechanics eliminates diamagnetic behavior.\cite{Griffiths-di}%
\ \ Within classical physics, single-particle diamagnetism, just like the
Planck spectrum, depends upon the inclusion of classical electromagnetic
zero-point radiation,\cite{Marshall} something which is incompatible with
nonrelativistic classical statistical mechanics. \ Again, our discussion,
which includes classical zero-point radiation, will depend upon simple
nonrelativistic systems which can be regarded as the low-velocity limits of
relativistic systems. \ 

\subsubsection{Isotropic Dipole Oscillator in a Magnetic Field}

In order to discuss diamagnetism, we consider a three-dimensional
harmonic-oscillator potential $V(\mathbf{r})=(1/2)\kappa r^{2}=(1/2)\kappa
(x^{2}+y^{2}+z^{2})$ in which there is a nonrelativistic particle of charge
$e$ and mass $m$, in the presence of a uniform magnetic field $\mathbf{B}$
along the $z$-direction $\mathbf{B}=\widehat{k}B$. \ Taking $\omega
_{0}=(\kappa/m)^{1/2},$ the nonrelativistic equation of motion for the
particle is
\begin{equation}
m\mathbf{\ddot{r}}=-m\omega_{0}^{2}\mathbf{r}+e(\mathbf{\dot{r}/c)\times
B}+m\tau\mathbf{\dddot{r}}+e\mathbf{E}_{T}(0,t) \label{SHO3}%
\end{equation}
where $-m\omega_{0}^{2}\mathbf{r}$ is the force due to the harmonic-oscillator
potential, $e(\mathbf{\dot{r}/c)\times B}$ is the Lorentz force of the
magnetic field, $m\tau\mathbf{\dddot{r}}$ is the radiation damping force, and
$e\mathbf{E}_{T}(0,t)$ is the driving force due to the electric field of the
thermal radiation taken in the dipole approximation. \ After dividing through
by the mass $m,$ the vector equation (\ref{SHO3}) can be rewritten as three
component equations%
\begin{equation}
\ddot{x}=-\omega_{0}^{2}x+2\omega_{L}\dot{y}+\tau\dddot{x}+(e/m)E_{Tx}
\label{Eqa}%
\end{equation}%
\begin{equation}
\ddot{y}=-\omega_{0}^{2}y-2\omega_{L}\dot{x}+\tau\dddot{y}+(e/m)E_{Ty}
\label{Eqb}%
\end{equation}%
\begin{equation}
\ddot{z}=-\omega_{0}^{2}z+\tau\dddot{z}+(e/m)E_{Tz} \label{Eqc}%
\end{equation}
where
\begin{equation}
\omega_{L}=\omega_{B}/2=eB/(2mc). \label{Lar}%
\end{equation}
Here we have a system of three linear differential equations. \ The
steady-state solution for equation (\ref{Eqc}) was given earlier in Eq.
(\ref{Xt}). \ The first two equations (\ref{Eqa})\ and (\ref{Eqb}) are coupled
linear differential equations with steady-state solutions%
\begin{equation}
x=%
{\textstyle\sum_{\mathbf{k}}}
{\textstyle\sum_{\lambda=1}^{2}}
\frac{e}{m}\left(  \frac{2\pi U(\omega,T)}{V}\right)  ^{1/2}\left\{
\frac{(C\epsilon_{x}-i2\omega\omega_{L}\epsilon_{y})\exp[-i\omega
t+i\theta(\mathbf{k},\lambda)]}{C^{2}-(2\omega\omega_{L})^{2}}+cc\right\}
\label{XB}%
\end{equation}%
\begin{equation}
y=%
{\textstyle\sum_{\mathbf{k}}}
{\textstyle\sum_{\lambda=1}^{2}}
\frac{e}{m}\left(  \frac{2\pi U(\omega,T)}{V}\right)  ^{1/2}\left\{
\frac{(C\epsilon_{y}+i2\omega\omega_{L}\epsilon_{x})\exp[-i\omega
t+i\theta(\mathbf{k},\lambda)]}{C^{2}-(2\omega\omega_{L})^{2}}+cc\right\}
\label{YB}%
\end{equation}
where%
\begin{equation}
C=-\omega^{2}+\omega_{0}^{2}-i\tau\omega^{3} \label{C}%
\end{equation}

\subsubsection{System Magnetic Moment\ \ }

Because we cannot apply nonrelativistic classical statistical mechanics to a
system where zero-point energy is involved, we will consider not the system
energy but rather the magnetic moment $\mathbf{M}_{dia}$ of our diamagnetic
system; \ later we will compare this diamagnetic magnetic moment with that of
a different (paramagnetic) system where nonrelativistic classical statistical
mechanics can be legitimately applied. \ Symmetry for our diamagnetic system
dictates that only a $z$-component is possible for the magnetic moment. \ The
particle angular momentum $\mathbf{L}$\ has $L_{z}=m(x\dot{y}-y\dot{x})$ so
that from $\mathbf{M}=[e/(2mc)]\mathbf{L,}$ we have\cite{Jackson187}
\begin{equation}
\left\langle M_{z-dia}\right\rangle =\frac{e}{2mc}\left\langle L_{z}%
\right\rangle =\frac{e}{2c}\left\langle x\dot{y}-y\dot{x}\right\rangle .
\label{M1}%
\end{equation}
We can differentiate to obtain the time derivatives and the take the averages
in either time or over the random phases as in Eqs. (\ref{av1}) and
(\ref{av2}) so as to obtain%

\begin{equation}
\left\langle x\dot{y}\right\rangle =\frac{e^{2}}{m^{2}}%
{\textstyle\sum_{\mathbf{k}}}
{\textstyle\sum_{\lambda=1}^{2}}
\left(  \frac{2\pi U(\omega,T)}{V}\right)  \frac{(\epsilon_{x}^{2}%
+\epsilon_{y}^{2})(2\omega^{2}\omega_{L})[2(-\omega^{2}+\omega_{0}^{2}%
)]}{|\Lambda_{+}|^{2}|\Lambda_{-}|^{2}} \label{XYd1}%
\end{equation}
where%
\begin{equation}
\left\vert \Lambda_{+}\right\vert ^{2}=(-\omega^{2}+\omega_{0}^{2}%
+2\omega\omega_{L})^{2}+(\tau\omega^{3})^{2} \label{Lp}%
\end{equation}
and%
\begin{equation}
\left\vert \Lambda_{-}\right\vert ^{2}=(-\omega^{2}+\omega_{0}^{2}%
-2\omega\omega_{L})^{2}+(\tau\omega^{3})^{2} \label{Lm}%
\end{equation}

Following the pattern taking us from Eq. (\ref{Eav}) to Eq.(\ref{Eav2}), we
assume that the normal modes are closely spaced; we replace the summation over
$\mathbf{k}$ by an integral, integrate over angles, and sum over polarizations
to obtain%
\begin{align}
\left\langle x\dot{y}\right\rangle  &  =\frac{e^{2}}{m^{2}}\left(  \frac
{a}{2\pi}\right)  ^{3}%
{\textstyle\int_{\omega=0}^{\infty}}
\frac{d\omega}{c^{3}}\omega^{2}2\left(  \frac{2}{3}\right)  4\pi\left(
\frac{2\pi U(\omega,T)}{V}\right)  \frac{(2\omega^{2}\omega_{L})[2(-\omega
^{2}+\omega_{0}^{2})]}{|\Lambda_{+}|^{2}|\Lambda_{-}|^{2}}\nonumber\\
&  =\frac{e^{2}}{m^{2}}%
{\textstyle\int_{\omega=0}^{\infty}}
\frac{d\omega}{c^{3}}\omega^{2}\frac{8}{3\pi}U(\omega,T)\frac{(2\omega
^{2}\omega_{L})(-\omega^{2}+\omega_{0}^{2})}{|\Lambda_{+}|^{2}|\Lambda
_{-}|^{2}} \label{XYd2}%
\end{align}
Now for positive $\omega,$ the quantity $\left\vert \Lambda_{+}\right\vert
^{2}$ takes its minimum value when $-\omega^{2}+\omega_{0}^{2}+2\omega
\omega_{L}=0$ or
\begin{equation}
\omega=\omega_{+}=(\omega_{0}^{2}+\omega_{L}^{2})^{1/2}+\omega_{L} \label{Wp}%
\end{equation}
and the quantity $\left\vert \Lambda_{-}\right\vert ^{2}$ takes its minimum
value when $-\omega^{2}+\omega_{0}^{2}-2\omega\omega_{L}=0$ or%
\begin{equation}
\omega=\omega_{-}=(\omega_{0}^{2}+\omega_{L}^{2})^{1/2}-\omega_{L} \label{Wm}%
\end{equation}
\ If the quantities $\tau\omega_{-}$ and $\tau\omega_{+}$ are small,
$\tau\omega_{-}<<1,~\tau\omega_{+}<<1,$ corresponding to small radiation
damping, the integral in Eq. (\ref{XYd2}) is sharply peaked at $\omega_{+}$
and $\omega_{-}.$ \ Thus we will evaluate the integral in the approximation of
two resonances, one at $\omega_{+}$ and one at $\omega_{-}$. \ We replace
every appearance of the frequency $\omega$ by $\omega_{+}$ or by $\omega_{-},$
except in $\left\vert \Lambda_{+}\right\vert ^{2}$\ and $\left\vert
\Lambda_{-}\right\vert ^{2}$\ where the combination $\omega-\omega_{+}$ or
$\omega-\omega_{-}$ appears, so that%
\begin{equation}
\left\vert \Lambda_{+}\right\vert ^{2}\approx4(\omega_{0}^{2}+\omega_{L}%
^{2})(\omega-\omega_{+})^{2}+(\tau\omega_{+}^{3})^{2} \label{Lp2}%
\end{equation}
and \
\begin{equation}
\left\vert \Lambda_{-}\right\vert ^{2}\approx4(\omega_{0}^{2}+\omega_{L}%
^{2})(\omega-\omega_{-})^{2}+(\tau\omega_{-}^{3})^{2} \label{Lm2}%
\end{equation}
Now treating each resonant term separately, we extend the integrals over
$\omega$ from $-\infty$ to $+\infty$ to obtain%
\begin{align}
\left\langle x\dot{y}\right\rangle  &  =%
{\textstyle\int_{-\infty}^{\infty}}
d\omega\,\frac{\omega_{+}^{2}}{c^{3}}\frac{e^{2}}{m^{2}}\frac{8}{3\pi}%
U(\omega_{+},T)\frac{(2\omega_{+}^{2}\omega_{L})(-\omega_{+}^{2}+\omega
_{0}^{2})}{[4(\omega_{0}^{2}+\omega_{L}^{2})(\omega-\omega_{+})^{2}%
+(\tau\omega_{+}^{3})^{2}][-4\omega_{L}\omega_{+}]^{2}}\nonumber\\
&  +%
{\textstyle\int_{-\infty}^{\infty}}
d\omega\frac{\omega_{-}^{2}}{c^{3}}\frac{e^{2}}{m^{2}}\frac{8}{3\pi}%
U(\omega_{-},T)\frac{(2\omega_{-}^{2}\omega_{L})(-\omega_{-}^{2}+\omega
_{0}^{2})}{[4\omega_{L}\omega_{-}]^{2}[4(\omega_{0}^{2}+\omega_{L}^{2}%
)(\omega-\omega_{-})^{2}+(\tau\omega_{-}^{3})^{2}]} \label{XYd3}%
\end{align}
Using the integral in Eq. (\ref{Ib}), we have%
\begin{align}
\left\langle x\dot{y}\right\rangle  &  =\frac{e^{2}}{m^{2}}\frac{\omega
_{+}^{2}}{c^{3}}\frac{8}{3\pi}U(\omega_{+},T)\frac{(2\omega_{+}^{2}\omega
_{L})(-\omega_{+}^{2}+\omega_{0}^{2})}{[-4\omega_{L}\omega_{+}]^{2}}\frac{\pi
}{2(\omega_{0}^{2}+\omega_{L}^{2})^{1/2}(\tau\omega_{+}^{3})}\nonumber\\
+  &  \frac{e^{2}}{m^{2}}\frac{\omega_{-}^{2}}{c^{3}}\frac{8}{3\pi}%
U(\omega_{-},T)\frac{(2\omega_{-}^{2}\omega_{L})(-\omega_{-}^{2}+\omega
_{0}^{2})}{[4\omega_{L}\omega_{-}]^{2}}\frac{\pi}{2(\omega_{0}^{2}+\omega
_{L}^{2})^{1/2}(\tau\omega_{-}^{3})}\nonumber\\
&  =-\frac{1}{2m}\frac{U(\omega_{+},T)}{(\omega_{0}^{2}+\omega_{L}^{2})^{1/2}%
}+\frac{1}{2m}\frac{U(\omega_{-},T)}{(\omega_{0}^{2}+\omega_{L}^{2})^{1/2}}
\label{XYd4}%
\end{align}
where we have noted that $(-\omega_{+}^{2}+\omega_{0}^{2})=-2\omega_{L}%
\omega_{+}$ and $(-\omega_{-}^{2}+\omega_{0}^{2})=+2\omega_{L}\omega_{-}$ We
can evaluate the average $\left\langle -y\dot{x}\right\rangle $ in a similar
fashion and find that it is equal to $\left\langle x\dot{y}\right\rangle .$ \ 

\subsubsection{Result for the Magnetic Moment - Single-Particle Diamagnetism}

Thus combining $\left\langle x\dot{y}\right\rangle $ and $\left\langle
-y\dot{x}\right\rangle ,$ we find that the magnetic moment in the presence of
a magnetic field is given by%
\begin{equation}
\left\langle M_{z-dia}\right\rangle =\frac{e}{2c}\left\langle x\dot{y}%
-y\dot{x}\right\rangle =-\frac{e}{2mc(\omega_{0}^{2}+\omega_{L}^{2})^{1/2}%
}\left[  U(\omega_{+},T)-U(\omega_{-},T)\right]  \label{M3}%
\end{equation}

In our analysis thus far, we have arrived at only the asymptotic limits for
the thermal radiation energy $U(\omega,T)$, which are given in Eqs.
(\ref{Uh1}) and (\ref{zpr}). \ In the high-temperature limit where
$k_{B}T>>\hbar\omega$ for all frequencies of interest, we recover the results
of nonrelativistic classical statistical mechanics where the energy
$U(\omega,T)$ becomes the energy equipartition value $U$($\omega,T)\rightarrow
k_{\mathbf{B}}T$ for every frequency, so that the magnetic moment expression
in Eq. (\ref{M3}) vanishes
\begin{equation}
\left\langle M_{z-dia}\right\rangle \rightarrow-\frac{e}{2mc(\omega_{0}%
^{2}+\omega_{L}^{2})^{1/2}}\left[  k_{\mathbf{B}}T-k_{\mathbf{B}}T\right]
=0\text{ \ for energy equipartition} \label{M4}%
\end{equation}
This result agrees with the Bohr-van Leuwen theorem for the absence of
diamagnetism in nonrelativistic classical statistical mechanics.\cite{BVL}
$\ $

On the other hand, the low temperature limit indeed shows diamagnetic behavior
arising from zero-point radiation since $U(\omega,0)=(1/2)\hbar\omega)$. \ We
find that in this low-temperature limit Eq. (\ref{M3}) becomes
\begin{align}
\left\langle M_{z-dia}\right\rangle  &  \rightarrow-\frac{e}{2mc(\omega
_{0}^{2}+\omega_{L}^{2})^{1/2}}\left[  \frac{1}{2}\hbar\omega_{+}-\frac{1}%
{2}\hbar\omega_{-}\right] \nonumber\\
&  =-\frac{e\hbar\omega_{L}}{2mc(\omega_{0}^{2}+\omega_{L}^{2})^{1/2}%
}=-\left(  \frac{e}{2mc}\right)  ^{2}\frac{\hbar B}{(\omega_{0}^{2}+\omega
_{L}^{2})^{1/2}}\text{ \ for \ }T=0 \label{M5}%
\end{align}
where we have inserted $\omega_{L}=eB/(2mc)$ in the numerator. \ Thus we find
that the average angular momentum and magnetic moment do not vanish in the low
temperature limit. \ Single-particle diamagnetism as an equilibrium
thermodynamic property within classical physics depends on the existence of
Lorentz-invariant classical zero-point radiation.

\subsubsection{Free-Particle Diamagnetism}

The diamagnetic behavior of our system at zero temperature becomes even more
striking if we take the confining harmonic oscillator potential as extremely
weak, $\omega_{0}\rightarrow0.$ \ In this case, equation (\ref{M5}) gives the
system magnetic moment%
\begin{equation}
\left\langle M_{z-dia}\right\rangle =-\frac{e\hbar\omega_{L}}{2mc(\omega
_{0}^{2}+\omega_{L}^{2})^{1/2}}\rightarrow\frac{e\omega_{L}}{2mc|\omega_{L}%
|}\hbar=-\frac{|e|\hbar}{2mc}\text{ \ for }\omega_{0}\rightarrow0,\text{
}T\rightarrow0 \label{M6}%
\end{equation}
The absolute values arises because the direction of rotation $\omega
_{B}=eB/(mc)$ involved in cyclotron motion reverses sign with the sign of the
charge. \ The magnetic moment is always such as to give diamagnetic behavior.

It is also curious to see that the angular momentum $\left\langle
L_{z}\right\rangle =(2mc/e)\left\langle M_{z}\right\rangle $ of the free
charge takes the values
\begin{equation}
\left\langle L_{z}\right\rangle =-\frac{e}{|e|}\hbar=\mp\hbar\label{Lh}%
\end{equation}
where the sign of the charge determines the plus or minus sign and the
direction of the angular momentum is determined by the direction of the
magnetic field.\cite{SS}

\section{Derivation of the Planck Spectrum}

\subsection{Use of the Diamagnetic System}

\subsubsection{Large-Mass Limit for the Diamagnetic System}

In our analysis thus far, the values for the average thermal energy per normal
mode $U(\omega,T)$ and magnetic moment $\left\langle M_{z}\right\rangle $ at a
general temperature $T$ are unknown because the spectrum of thermal radiation
has not yet been obtained. \ So far, we know only the asymptotic
high-temperature and low-temperature limits. \ In order to obtain the full
Planck spectrum within classical physics, we will make use of nonrelativistic
classical statistical mechanics applied in the appropriate limits.

First of all, at finite temperature $T,$ we go to the nonrelativistic limit
for our charged particle in the harmonic potential in a magnetic field. \ Thus
we can take the mass $m$ of our charge as very large so that the particle
velocity is small, $v/c<<1.$ \ Then the frequency $\omega_{0}=(\kappa
/m)^{1/2}$ associated with the confining potential becomes very small so that
$k_{\mathbf{B}}T>>\hbar\omega_{0}$, and we find the energy equipartition
behavior associated with all frequencies, except $\omega_{B}=eB/(mc)=2\omega
_{L}$ which we hold constant for by increasing the magnetic field $B$ to
offset the large mass $m$. \ In this limit where $\omega_{+}\rightarrow
\omega_{B}$ and $\omega_{-}\rightarrow0$, we find the magnetic moment at
finite temperature in Eq. (\ref{M3}) becomes%
\begin{equation}
\left\langle M_{z-dia}\right\rangle =-\frac{e}{mc\omega_{B}}\left[
U(\omega_{B},T)-k_{\mathbf{B}}T\right]  \label{MT}%
\end{equation}
where the crucial energy function $U(\omega_{B},T)$ corresponding to the
average energy of a normal mode at frequency $\omega_{B}$ and temperature $T$
is still unknown. \ 

\subsubsection{Paramagnetic Behavior at Temperature $T$}

In order to obtain $U(\omega_{B},T)$, we will compare paramagnetic and
diamagnetic behavior at temperature $T$. \ We consider a three-dimensional
paramagnetic rotator of magnetic moment $\mu$ and moment of inertia $I$ which
rotates in the presence of a magnetic field $\mathbf{B}$ along the $z$-axis.
\ In the limit that the moment of inertial $I$ is taken very large
$I\rightarrow\infty$, the rotator system will be that of a free
nonrelativistic rotator, rotating at arbitrarily low frequency so that the
zero-point radiation will make no contribution to the energy. \ Thus we can
apply nonrelativistic classical statistical mechanics. \ In this case, the
canonical phase space variables are the angular momentum $\mathbf{L}%
=I\overrightarrow{\omega}$ and the angles, $\theta,\phi.$ The partition
function associated with the kinetic energy of rotation can be treated
separately from that associated with the energy of orientation so
that\cite{Morse}%
\begin{equation}
Z_{\theta}=%
{\textstyle\int_{0}^{\pi}}
d\theta\sin\theta\exp\left[  -\frac{(-\mu B\cos\theta)}{k_{\mathbf{B}}%
T}\right]  =\left(  \frac{2k_{\mathbf{B}}T}{\mu B}\right)  \sinh\left(
\frac{\mu B}{k_{\mathbf{B}}T}\right)  \label{ZM}%
\end{equation}
The average magnetic moment $\left\langle M_{z-para}\right\rangle $\ is given
by%
\begin{equation}
\left\langle M_{z-para}\right\rangle =\left\langle \mu\cos\theta\right\rangle
=\frac{1}{B}\frac{\partial(\ln Z_{\theta})}{\partial\left(  1/k_{\mathbf{B}%
}T\right)  }=\mu\left[  \coth\left(  \mu B/k_{\mathbf{B}}T\right)  -\left(
k_{\mathbf{B}}T/\mu B\right)  \right]  \label{Mpara}%
\end{equation}
The hyperbolic cotangent function has the expansion for small argument $x<<1$
given by $\coth x=1/x+x/3-x^{3}/45+...,$ while for large $x>>1,$
$\coth(x)\rightarrow1.~\ $Therefore the magnetic moment $\left\langle
M_{z-para}\right\rangle $\ of this nonrelativistic paramagnetic system has the
asymptotic limits%
\begin{equation}
\left\langle M_{z-para}\right\rangle \rightarrow\mu\text{ \ for \ }%
T\rightarrow0\label{Mpara0}%
\end{equation}
and
\begin{equation}
\left\langle M_{z-para}\right\rangle \rightarrow0\text{ \ \ for \ }%
T\rightarrow\infty.\label{MparaT}%
\end{equation}

\subsubsection{Obtaining the Planck Spectrum}

The asymptotic limits of the magnetic moment for the nonrelativistic
paramagnetic rotator are analogous to those we have found for the diamagnetic
magnetic moment of our nonrelativistic free particle in a magnetic field.
\ Indeed, we can imagine a thermodynamic system consisting of our paramagnetic
rotator and our diamagnetic free particle in a uniform magnetic field and
taken sufficiently far apart that the magnetic interaction between them is
negligible. \ If we take the magnitude of the paramagnetic moment as
$\mu=e\hbar/(2mc),$ then, from Eqs. (\ref{MT}) and (\ref{Mpara}), the total
magnetic moment for the system is
\begin{align}
\left\langle M_{z-para}\right\rangle +\left\langle M_{z-dia}\right\rangle  &
=\frac{e\hbar}{2mc}\left[  \coth\left(  \frac{\hbar\omega_{B}}{2k_{\mathbf{B}%
}T}\right)  -\frac{2k_{\mathbf{B}}T}{\hbar\omega_{B}}\right]  -\frac{e}%
{mc}\left[  \frac{1}{\omega_{B}}U\left(  \frac{\omega_{B}}{k_{\mathbf{B}}%
T}\right)  -\frac{k_{\mathbf{B}}T}{\omega_{B}}\right] \nonumber\\
&  =\frac{e\hbar}{2mc}\left[  \coth\left(  \frac{\hbar\omega_{B}%
}{2k_{\mathbf{B}}T}\right)  -\frac{2}{\hbar\omega_{B}}U\left(  \frac
{\omega_{B}}{k_{\mathbf{B}}T}\right)  \right]  \label{Mtot}%
\end{align}
which has the asymptotic limits%
\begin{equation}
\left\langle M_{z-para}\right\rangle +\left\langle M_{z-dia}\right\rangle
\rightarrow0\text{ \ \ for \ \ }T\rightarrow0 \label{Mtot0}%
\end{equation}%
\begin{equation}
\left\langle M_{z-para}\right\rangle +\left\langle M_{z-dia}\right\rangle
\rightarrow0\text{ \ \ for \ \ }T\rightarrow\infty. \label{MtotT}%
\end{equation}
There is no hidden structure in our magnetic moment system so that the only
allowed interpolation between the two limits is the vanishing of the total
magnetic moment at all temperature. \ But then from the vanishing of the
second line of Eq. (\ref{Mtot}), we must have the energy per normal mode of
the blackbody radiation spectrum as
\begin{equation}
U(\omega,T)=\frac{1}{2}\hbar\omega_{B}\coth\left(  \frac{\hbar\omega
}{2k_{\mathbf{B}}T}\right)  =\frac{\hbar\omega}{\exp[\hbar\omega
/k_{\mathbf{B}}T]-1}+\frac{1}{2}\hbar\omega\label{Planck}%
\end{equation}
which is exactly the Planck spectrum. \ This spectrum agrees with both the old
thermal radiation measurements of Lummer and Pringheim and the recent Casimir
force measurements. \ If we insert this spectrum into the right-hand side of
Eq. (\ref{M3}) for the magnetic moment, we have the results of Landau
diamagnetism.\cite{Landau}

\section{Closing Summary}

In this article, we show that classical physics which includes classical
electromagnetic zero-point radiation and uses relativity appropriately indeed
predicts both the Planck spectrum of blackbody radiation and the presence of
Landau diamagnetism at thermodynamic equilibrium. \ Appropriate use of
relativity requires that nonrelativistic systems be considered only in the
limit of zero particle velocity. \ In addition, nonrelativistic statistical
mechanics can be applied only in the zero-velocity limit for situations where
the zero-point energy makes no contribution. \ 

The conclusions of this manuscript are unsettling to many physicists.
\ Although the basic analysis has been in the research literature for a number
of years, the ideas have never entered the textbook literature. \ The
existence of classical electromagnetic zero-point radiation and the importance
of relativity were ideas unfamiliar to the physicists of the early 20th
century. \ These ideas are still unfamiliar today.

\section{Acknowledgement}

I wish to thank Dr. Hanno Ess\'{e}n for his helpful comments on the Bohr-van
Leeuwen theorem, and for sending me a copy of his review of diamagnetic
behavior appearing in ref. 4.

\bigskip

\begin{thebibliography}{99}
\bibitem {Blanco}There is one scattering calculation which used the
relativistic expressions for the momentum of the scattering charged particle.
\ However, the calculation limited itself to nonrelativistic potential
functions which excluded the Coulomb potential and so was not a relativistic
calculation. \ See Blanco, Pesquera, and Santos in ref. 6.
\bibitem {BVL}See the discussion by Essen and Fiolhais in ref. 4. \ In the
large-mass limit for the analysis in this article, the magnetic-field-energy
terms will vanish, and we indeed expect to recover the results of the Bohr-van
Leeuwen theorem when there is no zero-point radiation.
\end{thebibliography}
\end{document}